\documentclass[aps,twocolumn,groupedaddress]{revtex4-2}

\usepackage{amsmath}
\usepackage{amsfonts}
\usepackage{amssymb}
	
\usepackage{braket}

\newcommand{\Tr}{\text{Tr}}

\begin{document}

\title{Optimal quantum channels}

\author{Rocco Duvenhage}
\email[]{rocco.duvenhage@up.ac.za}
\affiliation{Department of Physics, University of Pretoria, Pretoria, South Africa}

\date{September 12, 2021}

\begin{abstract}
A method to optimize the cost of a quantum channel is developed. The goal is to determine the cheapest channel that produces prescribed output states for a given set of input states. This is essentially a quantum version of optimal transport.
To attach a clear conceptual meaning to the cost, channels are viewed in terms of what we call elementary transitions, which are analogous to point-to-point transitions between classical systems. 
The role of entanglement in optimization of cost is emphasized. 
We also show how our approach can be applied to theoretically search for channels performing a prescribed set of tasks on the states of a system, while otherwise disturbing the state as little as possible.
\end{abstract}

\maketitle


\section{Introduction}

Quantum channels are ubiquitous in quantum information theory. A natural question is: what is the cost incurred when using a channel and how can it be optimized? It may lead to conceptual insights regarding quantum channels and can guide us in using resources most efficiently in applications.

One may want the cheapest channel (or channels) to perform a particular task. For example, to produce certain output states for a given set of input states, which is exactly the problem we study.

Previous literature focussed on the cost of the input states \cite{CHP, Jar, DPW}, for channel capacity per unit cost, whereas here we assign cost to the channel itself, in line with \cite{DePT}.

In outline our approach is as follows: Given two systems $A$ and $B$ with finite dimensional Hilbert spaces, we consider a set of states (density matrices) $\rho^A_1,...,\rho^A_l$ of the former, as well as a set $\rho^B_1,...,\rho^B_l$ of the latter, and require a channel $\mathcal{E}$ from $A$ to $B$ to satisfy
\begin{equation}
\mathcal{E}(\rho^A_j)=\rho^B_j
\label{rand}
\end{equation}
for $j=1,...,l$. Using the Choi-Jamio{\l}kowski duality \cite{deP, J, C} between channels and states, we represent $\mathcal{E}$ as a density matrix $\kappa_\mathcal{E}$, and express its cost as
\begin{equation}
K_C(\mathcal{E})=\Tr(C\kappa_\mathcal{E}),
\label{koste}
\end{equation}
where the \emph{cost matrix} $C$ is an observable of the composite system $AB$. This cost $K_C(\mathcal{E})$ is the expected cost of what will be referred to as ``elementary transitions" in the channel.

One then finds the optimal channel or channels, by minimizing the cost while satisfying the conditions (\ref{rand}).

This is closely related to quantum optimal transport, which corresponds to the case $l=1$. The approaches to quantum optimal transport closest to our outline above, appears in \cite{DePT, D}. Other approaches appeared in \cite{ZS, CM1, CM2, CM3, YYGT, CGT, CGGT, GMP, GP, CGP1, CGP2, AF, PCVS, DR, Ik, dePMTL, KdePMLL, FECZ, CEFZ}. 

There may be some terminological confusion here: ``Quantum optimal transport" is a quantum version of optimal transport of probability \cite{V}, as opposed to ``quantum transport" which refers to current flow in a quantum context, for example of electrons \cite{Dat}. 

A key difference between this paper and many of the references mentioned above, including \cite{DePT}, is that here we do not focus on quantum Wasserstein distances. We nevertheless use a result regarding quadratic Wasserstein distances from \cite{D} in Section \ref{AfdMinVer}. 

We note that \cite{DePT} does use the correspondence between states and channels, and expresses the transport cost in the same way as above. But, in their setup, the cost is in effect relative to a single input state. 

Here we rather set the cost up to be independent of any particular input state, by using the maximally mixed state as a reference state. This is more appropriate for our goals, as the cost matrix then applies to all input states $\rho^A_j$ and channels in (\ref{rand}). 

In addition, we attach a clearer conceptual meaning to the cost, by introducing the idea of elementary transitions as a quantum analogue of point to point transitions in classical spaces. 

We also emphasize the role of entanglement, which seems to have received limited attention in the literature on quantum optimal transport. See \cite{dePMTL} for a different aspect of entanglement, namely robustness against perturbations of the Wasserstein distance. We instead focus on how entanglement can directly contribute to lowering cost.

Our basic framework is set out in Sections \ref{AfdElemOorg} and \ref{AfdKoste}. This includes the introduction of elementary transitions.
The role of entanglement is discussed in Section \ref{AfdRolVanVerstr}. 
This is followed by two examples in Section \ref{AfdVbe}, illustrating a number of the ideas from the preceding sections. 
In Section \ref{AfdMinVer} we apply our setup to obtain channels performing tasks of the form (\ref{rand}), where $A$ and $B$ are the same system, while causing a minimal disturbance to other states of $A$, as indirectly measured by the cost matrix. Sections \ref{AfdVbe} and \ref{AfdMinVer} can be read independently of one another. Our concluding remarks appear in Section \ref{AfdAfsl}.

\section{Channels, states and elementary transitions}\label{AfdElemOorg}

To realize the outline above, we use the Choi-Jamio{\l}kowski duality (see the review \cite{JLF}). Fix an orthonormal basis $\ket{1^A},...,\ket{m^A}$ for the Hilbert space $H_A$ of the system $A$, and define a state $\kappa_\mathcal{E}$ of $AB$ by
\begin{equation}
\kappa_\mathcal{E}=\frac{1}{m}\sum_{ij}\ket{i^A}\bra{j^A}\otimes\mathcal{E}(\ket{i^A}\bra{j^A}).
\label{CJ}
\end{equation}
This state reduces to the maximally mixed state of $A$, so the latter state serves as a reference state in our setup. 

Diagonalize this state as 
\begin{equation}
\kappa_\mathcal{E}=\sum_{\alpha=1}^{mn}p_\alpha\kappa_\alpha
\label{diagE}
\end{equation}
where $n$ is the dimension of the Hilbert space $H_B$ of $B$, the $\kappa_\alpha$ are pure states and the $p_\alpha$ are probabilities. We aim to view these pure states as representing ``elementary transitions" which constitute the channel $\mathcal{E}$. 

Invert the duality to define linear maps $\mathcal{E}_\alpha$ from the space $L(H_A)$ of linear operators on $H_A$, to the space $L(H_B)$:
\[
\mathcal{E}_\alpha(\ket{i^A}\bra{j^A})=
m(\bra{j^A}\otimes I_B) 
\kappa_\alpha^{PT} 
(\ket{i^A}\otimes I_B),
\]
where $PT$ denotes the partial transpose with respect to the basis $\ket{1^A},...,\ket{m^A}$ for $H_A$, and $I_B$ is the identity operator on $H_B$. These maps are completely positive by Choi's method \cite{C}, but they are not necessarily channels.

From (\ref{CJ}) we have
\[
\mathcal{E}(\ket{i^A}\bra{j^A})=
m(\bra{j^A}\otimes I_B) 
\kappa_\mathcal{E}^{PT} 
(\ket{i^A}\otimes I_B),
\]
hence
\begin{equation}
\mathcal{E}=\sum_{\alpha}p_\alpha\mathcal{E}_\alpha.
\label{ontb}
\end{equation}
That is, we have decomposed the channel $\mathcal{E}$ into the completely positive maps $\mathcal{E}_\alpha$. Only the $\mathcal{E}_\alpha$ with $p_\alpha>0$ are viewed as part of the decomposition. This is a variation on a standard decomposition of a channel (see \cite{AP}), now specifically having a correspondence between the maps $\mathcal{E}_\alpha$ and the pure states $\kappa_\alpha$. As the diagonalization of $\kappa_\mathcal{E}$ is in general not unique (one can choose different orthonormal bases in an eigenspace of dimension greater than one), this decomposition of a channel is also not in general unique.

We call any $\mathcal{E}_\alpha$ appearing in a decomposition (\ref{ontb}) of a channel, an \emph{elementary transition}. 
It is analogous to classical transport from a point $x$ in one classical probability space, to a point $y$ in another, as in optimal transport. 

To clarify this analogy, we briefly describe what happens in classical optimal transport (refer to \cite{V} for an overview of classical optimal transport): 

The problem is to find the cheapest way to transport goods from one set of points to another. For example, from warehouses to shops. A cost $c(x,y)$ is incurred when moving one unit, say a truckload, of the goods from warehouse $x$ to shop $y$. The exact route from $x$ to $y$ is not relevant in this setup and is assumed to be determined by $x$ and $y$. Only the pair $(x,y)$ consisting of the initial point $x$ and the end point $y$ is needed. Therefore, in this picture, one truckload of goods transported from $x$ to $y$, is viewed as an elementary transition, denoted by $(x,y)$. Then $c(x,y)$ is seen as the cost of this elementary transition.

Mathematically one often models the distributions of goods over warehouses and shops, respectively, as probability distributions over each of the two sets, i.e., the total available goods is normalized to $1$. It is this probabilistic viewpoint which is analogous to the quantum setting.

Note, in particular, that the quantum pure state $\kappa_\alpha$ corresponding to $\mathcal{E}_\alpha$ is analogous to the pure state $(x,y)$ of a classical composite system. In this way an elementary transition $\mathcal{E}_\alpha$, via its duality with $\kappa_\alpha$, is analogous to a classical elementary transition $(x,y)$.

The cost associated to an elementary transition $\mathcal{E}_\alpha$ will be discussed in the next section.

The quantum case allows for much more interesting elementary transitions than the classical case when $\kappa_\alpha$ is entangled. The elementary transitions are then essentially non-classical. For example, any channel from $A$ to itself given by a unitary operator $U$ on $H_A$,
\[
\mathcal{E}(\rho^A)=U\rho^A U^\dagger,
\]
is an elementary transition dual to the maximally entangled pure state  $\kappa_\mathcal{E}$. Classical transport, on the other hand, just allows point to point elementary transitions. There are no classical elementary transitions involving larger portions (or the whole) of the probability spaces involved: the pairs $(x,y)$ are the only pure states of a classical composite system, i.e., of the Cartesian product of the two classical probability spaces.

One can refine this picture, and view an elementary transition in effect as a map from a subset of the set of states of $A$, to its image as a subset of $B$'s states. This can be seen by studying the \emph{support} of such a transition $\mathcal{E}_\alpha$, i.e., the Hilbert subspace $H_A^\alpha$ of $H_A$, orthogonal to the set of state vectors $\ket{\psi^A}$ in $H_A$ such that $\mathcal{E}_\alpha(\ket{\psi^A}\bra{\psi^A})=0$.
Unlike a channel $\mathcal{E}$, the support of which is always the whole of $H_A$ (as channels preserve the trace), the support of an elementary transition can be smaller.
Let $\mathcal{S}_A^\alpha$ be the set of density matrices on $H_A$ which are direct sums of density matrices on $H_A^\alpha$ and zero matrices on its orthogonal complement (i.e., arranged diagonally as two blocks).
By restricting an elementary transition $\mathcal{E}_\alpha$ to $\mathcal{S}_A^\alpha$, so in effect to density matrices on $H_A^\alpha$, one has a more refined representation $\mathcal{E}_\alpha'$ of the elementary transition, mapping from $\mathcal{S}_A^\alpha$ to its image $\mathcal{E}_\alpha(\mathcal{S}_A^\alpha)$. Such a restriction is natural exactly because $\mathcal{E}_\alpha$ takes states $\ket{\psi^A}\bra{\psi^A}$, with $\ket{\psi^A}$ orthogonal to $H_A^\alpha$, to zero, and can give some insight into the nature of an elementary transition. But, even rescaling this restricted map by scalar multiplication, in general still does not make it a channel, as will be seen by example near the end of Section \ref{AfdVbe}. This refined picture will not be used in this paper, though.

We note that every pure state of the composite system $AB$, corresponds to an elementary transition, in other words, it appears in a diagonalization of the form (\ref{diagE}) for some channel $\mathcal{E}$. In fact, the channel given by
\[
\mathcal{E}(X)=\frac{1}{n}\Tr(X)I_n,
\]
for any $m\times m$ matrix $X$, leads to
\[
\kappa_\mathcal{E}=\frac{1}{mn}I_{mn},
\]
the diagonalization  (\ref{diagE}) of which can be chosen to include any pure state of $AB$. 

In particular, we can view an elementary transition $\varepsilon$ as an independent object, without reference to a channel. Such an $\varepsilon$ is defined as the dual of any pure state $\rho^{AB}_{\text{pure}}$ of $AB$, via the inverse of (\ref{CJ}):
\[
\varepsilon(\ket{i^A}\bra{j^A})=
m(\bra{j^A}\otimes I_B) 
\left(  \rho^{AB}_{\text{pure}} \right) ^{PT} 
(\ket{i^A}\otimes I_B).
\]

We'll correspondingly occasionally refer to a pure state of $AB$ as an elementary transition.

\section{The cost matrix} \label{AfdKoste}

The cost associated to a channel will be encoded by a self-adjoint operator $C$ from $H_A\otimes H_B$ to itself, which will be referred to as the \emph{cost matrix}. 

The cost (\ref{koste}) of a channel can be viewed as the expected value of the costs $K_C(\mathcal{E_\alpha})=\Tr(C\kappa_\alpha)$ of the elementary transitions appearing in a decomposition (\ref{ontb}) of the channel. 

The cost matrix is analogous to the cost function $c(x,y)$ appearing in classical transport, which represents the cost of an elementary transition from point $x$ to point $y$. The cost (\ref{koste}) is analogous to the cost in classical optimal transport, given by the integral of $c(x,y)$ with respect to a measure (roughly analogous to $\kappa_\mathcal{E}$) who's marginals are the initial and final probability measures respectively. Here our conceptual setup diverges somewhat from classical optimal transport, as we allow a set of initial and final states, rather than just one of each. This is why we set up $\kappa_\mathcal{E}$ to always reduce to the maximally mixed state of $A$, rather than to a specific input state.

One way of representing or constructing $C$ is
\begin{equation}
C=\sum_{\alpha}k_\alpha\rho_\alpha^{AB},
\label{Cvoorst}
\end{equation}
for any finite set of pure states $\rho_\alpha^{AB}$ of $AB$, not necessarily orthogonal on Hilbert space level. This assigns the cost $k_\alpha$ (any real number) to the pure state $\rho_\alpha^{AB}$ representing an elementary transition, in analogy to the classical cost function,  where the cost $c(x,y)$ is assigned to the classical pure state $(x,y)$ representing the elementary transition from $x$ to $y$. 

Allowing non-orthogonal states, is non-classical. Indeed, cost matrices can be constructed by (\ref{Cvoorst}) with the states $\rho_\alpha^{AB}$ not orthogonal, allowing the eigenvalues of $C$ and even the optimal cost $K_C(\mathcal{E})$ to be lower than any of the constitutive costs $k_\alpha$. 
This is a straightforward but nevertheless decisive deviation from the classical case. 

When constructing $C$ using (\ref{Cvoorst}), we should avoid \emph{inadvertently} assigning zero cost to an elementary transition whose corresponding Hilbert space state vector is orthogonal to the states appearing in (\ref{Cvoorst}). We need to include at least $mn$ pure states $\rho_\alpha^{AB}$ whose Hilbert space vectors span the whole of $H_A\otimes H_B$. Including more than $mn$ states is analogous to, but more involved than, including more than one cost for a single point $(x,y)$ in the classical case, where such costs would simply be added together.

Expressions of the form
\begin{equation}
I_A\otimes O_B - O_A^T\otimes I_B,
\label{transp}
\end{equation}
are also useful building blocks for $C$,
where $O_A$ and $O_B$ are corresponding observables of $A$ and $B$ (say energy), measuring a difference in this observable. 
The transposition of $O_A$, with respect to the same basis used in the Choi-Jamio{\l}kowski duality, is natural due to general mathematical considerations regarding channel-state duality (see Section 3 and 7 of \cite{DS}, as well as \cite{DePT}).
This essentially reflects a dependence of the Choi-Jamio{\l}kowski duality on the maximally entangled state 
\begin{equation}
\ket{\Omega}=
\frac{1}{\sqrt{m}}\sum_{i=1}^{m}\ket{i^A}\ket{i^A},
\label{Omega}
\end{equation}
which is used in the duality (also see \cite{JLF}).

Both (\ref{Cvoorst}) and (\ref{transp}) will be illustrated in Sections \ref{AfdVbe} and \ref{AfdMinVer}, which then also serve to motivate (\ref{transp}).

\section{The role of entanglement}
\label{AfdRolVanVerstr}

Entanglement has an important role in optimizing cost. In short: 
An elementary transition is a channel only when the dual pure state is maximally entangled. Hence we can expect the optimal cost to be closer to a low eigenvalue of $C$, if some eigenvector corresponding to it is entangled. 
If no such eigenvector is maximally entangled, then none of them corresponds to a channel, requiring other elementary transitions to be included in order to build up a channel. The cost is consequently an expectation value including possibly higher eigenvalues, in turn leading to higher optimal cost of the channel.

To expand on this, keep in mind that if $c$ denotes the lowest eigenvalue of $C$, and $R$ is the corresponding eigenspace, then for all states $\rho^{AB}$ of $AB$, 
\[
\Tr(C\rho^{AB})\geq c,
\]
where $\Tr(C\rho^{AB})=c$ exactly for states $\rho^{AB}$ such that the image of $H_A\otimes H_B$ under $\rho^{AB}$ is contained in $R$, for example for states given by eigenvectors corresponding to $c$.

Also recall that (\ref{CJ}) gives a one-to-one correspondence between all channels (from $A$ to $B$) and the set of states of $AB$ which reduce to the maximally mixed state of $A$. The maximally entangled pure states of $AB$ are exactly the pure states reducing to the maximally mixed state of $A$, implying that an elementary transition $\varepsilon$ is a channel exactly when its dual
\[
\kappa_\varepsilon=
\frac{1}{m}\sum_{ij}\ket{i^A}\bra{j^A}\otimes\varepsilon(\ket{i^A}\bra{j^A})
\]
is a maximally entangled state of $AB$.

One consequence of these facts, for example, is that if $\dim(R)=1$ and the eigenvector $\ket{c}$ corresponding to $c$ is not maximally entangled, then 
\[
K_C(\mathcal{E})>c
\]
for all channels $\mathcal{E}$ from $A$ to $B$.

More generally, we now argue heuristically that entanglement in low-lying eigenvectors 
of $C$ (i.e., corresponding to low eigenvalues), tends
to lower the optimal cost.
Moreover, entanglement becomes more essential for low cost, the smaller the dimension of the low-lying eigenspaces.

To keep the cost of a channel as low as possible, 
we would like to ``build" it in the form (\ref{ontb}) from elementary transitions which are as close as possible to $C$'s low-lying eigenvectors 
and carrying probabilities $p_\alpha$ as large as possible.

If a pure state of $AB$ is far from being maximally entangled, then we can expect the dual elementary transition to be far from a channel. Consequently, we expect that in a decomposition (\ref{ontb}) of a channel, such an elementary transition will tend to carry a small probability.

On the other hand, a pure state $AB$ which is close to being maximally entangled, 
is dual to an elementary transition which is close to being a channel. 
Such an elementary transition has a greater chance to carry a large probability in a decomposition of a channel. 

Hence, the more entangled the low-lying eigenvectors of $C$ are, the 
better the chances that we can assign large probabilities to elementary transitions close to these eigenvectors, and still obtain a channel in the set of channels allowed by the requirements (\ref{rand}). As a result, the mentioned entanglement tends to lower the optimal cost.

This argument becomes more relevant the smaller the dimension of the eigenspaces of $C$ corresponding to lower eigenvalues. The larger the dimension of an eigenspace, the higher the chances of also being able to build a channel with the same cost, using non-entangled states 
in that eigenspace.

As will be seen in the next section, there are certainly limits to this heuristic argument, in particular with respect to how the size of the probabilities $p_\alpha$ are limited by the lack of entanglement of the associated pure states $\kappa_\alpha$. Nevertheless, it gives a strong indication that low optimal cost will tend to go hand in hand with high levels of entanglement in the low-lying eigenvectors of $C$.

A general class of cost matrices, which in absence of restrictions (\ref{rand}) leads to a unique maximally entangled state associated to optimal cost, with the identity channel as uniquely optimal, is discussed in Section \ref{AfdMinVer}.

\section{Examples in two dimensional Hilbert space}
\label{AfdVbe}

We consider two examples where the costs are respectively energy and time, to illustrate our setup in a simple context. In the process, elementary transitions are seen in action. Special attention is paid to the role of entanglement.

Assume that $H_A=H_B$ is two dimensional. We consider channels $\mathcal{E}$ from $A$ to itself. In terms of the notation 
\[
\rho=\rho^A,
\]
the general form of $\mathcal{E}$ is then
\[
\mathcal{E}(\rho)=\sum_{j=1}^{4}V_j\rho V_j^\dagger
\]
where 
\[V_j=
\left[ 
\begin{array}{cc}
a_j & b_j \\ 
c_j & d_j 
\end{array}
\right]
\]
are complex matrices satisfying 
$
\sum_{j}V_j^\dagger V_j=I_4
$
in the orthonormal basis we use for $H_A$. In terms of vectors $a$, $b$, $c$ and $d$ given by
\[a=(a_1,a_2,a_3,a_4)\] 
etc., and the usual complex dot product, we have
\begin{align*}
\mathcal{E}(\ket{1}\bra{1})= &
\left[ 
\begin{array}{cc}
a\cdot a & c\cdot a \\ 
a\cdot c & c\cdot c 
\end{array}
\right],\,
\mathcal{E}(\ket{1}\bra{2})= 
\left[ 
\begin{array}{cc}
b\cdot a & d\cdot a \\ 
b\cdot c & d\cdot c 
\end{array}
\right],
\\
\mathcal{E}(\ket{2}\bra{1})= &
\left[ 
\begin{array}{cc}
a\cdot b & c\cdot b \\ 
a\cdot d & c\cdot d 
\end{array}
\right],\,
\mathcal{E}(\ket{2}\bra{2})= 
\left[ 
\begin{array}{cc}
b\cdot b & d\cdot b \\ 
b\cdot d & d\cdot d 
\end{array}
\right].
\end{align*} 
Then $\kappa_\mathcal{E}$ is the $4\times 4$ matrix given by:
\[
\kappa_\mathcal{E}=\frac{1}{2}
\left[
\begin{array}{cc}
\mathcal{E}(\ket{1}\bra{1}) & \mathcal{E}(\ket{1}\bra{2}) \\ 
\mathcal{E}(\ket{2}\bra{1})  & \mathcal{E}(\ket{2}\bra{2}) 
\end{array}
\right].
\]

\subsection{Energy}

As first example,  assume that $A$ has the Hamiltonian 
\[
H=
\left[ 
\begin{array}{cc}
\epsilon/2 & 0 \\ 
0          & -\epsilon/2 
\end{array}
\right]
\]
with $\epsilon>0$.
Consider 
\[C=I_2\otimes H - H\otimes I_2+J\sigma_1\otimes\sigma_1\]
where $\sigma_1$ is the $x$ Pauli matrix. The first two terms in $C$ will tend to force $A$ to its lowest energy state ($z$ spin down), while the last term tends to preserve $x$ spin, with 
\[
J<0.
\] 
Then
\[
K_C(\mathcal{E})=
\frac{\epsilon}{2}(b\cdot b-c\cdot c)+
\frac{J}{2}(a\cdot d+d\cdot a+b\cdot c+c\cdot b),
\]
optimization of which requires $a$ to be proportional to $d$ and $b$ to $c$. 

The two negative eigenvalues of $C$ are  $J$ and $-\sqrt{J^2+\epsilon^2}$ with eigenvectors 
\[
\ket{\psi_1}=
\frac{1}{\sqrt{2}}(\ket{1}\ket{1}+\ket{2}\ket{2})
\]
and
\[
\ket{\psi_0}=
\frac{1}{\sqrt{L^2+J^2}}(L\ket{1}\ket{2}-J\ket{2}\ket{1})
\]
respectively, where $L=\sqrt{J^2+\epsilon^2}+\epsilon$. Note that as $\ket{\psi_0}$ is not maximally entangled, thus not dual to a channel, the lowest cost $-\sqrt{J^2+\epsilon^2}$ can not be reached by any channel.

As an example of a task the channel has to perform, we enforce the condition that the channel takes the lowest energy eigenstate to the highest, i.e.,
\begin{equation}
\mathcal{E}
\left[ 
\begin{array}{cc}
0 & 0 \\ 
0 & 1 
\end{array}
\right]
=
\left[ 
\begin{array}{cc}
1 & 0 \\ 
0 & 0 
\end{array}
\right].
\label{spinskuif}
\end{equation} 
This leads to the optimal cost
\[
K_C(\mathcal{E})=J,
\]
with the unique optimal channel $\mathcal{E}$ given by the unitary operator
\[ 
U=
\left[ 
\begin{array}{cc}
0 & 1 \\ 
1 & 0 
\end{array}
\right].
\]
This corresponds approximately to the entangled eigenvector $\ket{\psi_0}$. Even though the latter is not maximally entangled, it is close enough to the maximally entangled state vector 
\[
\ket{\psi_0'}=
\frac{1}{\sqrt{2}}(\ket{1}\ket{2}+\ket{2}\ket{1})
\]
(obtained when $J\rightarrow -\infty$) dual to $U$, that the latter channel is obtained as optimal.
This would not have been possible without entanglement of the eigenvector $\ket{\psi_0}$, which, combined with the fact that a maximally entangled elementary transition is a channel, lead to the optimal channel.

Now drop (\ref{spinskuif}). 
Writing $|b|=\sin\theta$ and $|c|=\sin\varphi$, we find the minimum of
\[
\frac{\epsilon}{2}(\sin^2\theta-\sin^2\varphi)+
J\cos(\theta-\varphi)
\]
over $\theta,\varphi\in[0,\pi/2]$, giving the optimal cost
\[
K_C(\mathcal{E})=
-\sqrt{J^2+\epsilon^2/4},
\]
with the optimal channels described by 
\[|a|^2=|b|^2=1-|c|^2=1-|d|^2=
\frac{1}{2}\left(1-\frac{\epsilon}{\sqrt{4J^2+\epsilon^2}}\right)
\]
for $a\cdot b+c\cdot d=0$, with $a$ proportional to $d$ and $b$ to $c$.
For $J\rightarrow 0$, the channel takes all states to the minimum energy eigenstate of $H$, as one would physically expect.

To clarify entanglement's role in this case, we study the limit 
\[J\rightarrow -\infty.
\] 
The optimal cost then asymptotically approaches both the negative eigenvalues of $C$, the optimal channels being
\begin{multline*}
\mathcal{E}
\left[ 
\begin{array}{cc}
\rho_{11} & \rho_{12} \\ 
\rho_{21} & \rho_{22} 
\end{array}
\right]
=\\
\frac{1}{2}\left[ 
\begin{array}{cc}
1-i\gamma(\rho_{12}-\rho_{21})                   & \rho_{12}+\rho_{21}-i\gamma(\rho_{11}-\rho_{22}) \\ 
\rho_{12}+\rho_{21}+i\gamma(\rho_{11}-\rho_{22}) & 1+i\gamma(\rho_{12}-\rho_{21})
\end{array}
\right]
\end{multline*}
for any $-1\leq\gamma\leq1$. The decomposition of such an $\mathcal{E}$ in elementary transitions, expressed in terms of states, is
\[
\kappa_\mathcal{E}=
\frac{1+\gamma}{2}\ket{\phi_1}\bra{\phi_1}+
\frac{1-\gamma}{2}\ket{\phi_2}\bra{\phi_2},
\]
where the elementary transitions
\[
\ket{\phi_1}=\frac{1}{\sqrt{2}}\ket{\psi_1}+\frac{i}{\sqrt{2}}\ket{\psi_0'}
\]
and
\[
\ket{\phi_2}=\frac{i}{\sqrt{2}}\ket{\psi_1}+\frac{1}{\sqrt{2}}\ket{\psi_0'}.
\]
are both maximally entangled.

The decomposition into the elementary transitions (expressed as state vectors) $\ket{\phi_1}$ and $\ket{\phi_2}$ above, is unique for $\gamma\neq0$, exactly because the two probabilities $(1+\gamma)/2$ and $(1-\gamma)/2$ differ. Hence, for $\gamma\neq0$, entangled elementary transitions are necessarily present in the corresponding optimal channels. 

For $\gamma=0$, on the other hand, the eigenspace of the single non-zero eigenvalue $1/2$ of $\kappa_\mathcal{E}$ becomes two-dimensional, and consequently non-entangled elementary transitions can also be obtained in the decomposition of the optimal channel given by $\gamma=0$.

For finite, but large enough $|J|$, the results above will still hold approximately, which means that highly entangled elementary transitions will necessarily carry large weight in most of the optimal channels.

In line with Section \ref{AfdRolVanVerstr}, we also see the following: Both the maximally entangled state vectors
$\ket{\psi_1}$ and $\ket{\psi_0'}$, respectively an eigenvector and approximate eigenvector of $C$ with lowest eigenvalues, appear as uniquely determined (when $\gamma\neq0$) components of the elementary transitions constituting the optimal channels, confirming that entanglement of the lower cost eigenvectors, aids in lowering optimal cost. 

\subsection{Time}

In our second example, we construct a cost matrix from four unitary channels given by 
\[
U_1=
\left[ 
\begin{array}{cc}
1 & 0 \\ 
0 & 1 
\end{array}
\right],
U_2=
\left[ 
\begin{array}{cc}
0 & 1 \\ 
1 & 0 
\end{array}
\right],
U_3=
\left[ 
\begin{array}{cc}
1 & 0 \\ 
0 & -1 
\end{array}
\right],
U_4=U_3 U_2
\]
calling the dual pure states $\kappa_1,...,\kappa_4$ respectively, and using (\ref{Cvoorst}):
\begin{equation}
C=k_1\kappa_1+...+k_4\kappa_4,
\label{kosteKonstr}
\end{equation}
with $k_j$ the cost of the elementary transition $U_j$.

Note that $k_1,...,k_4$ are the eigenvalues of $C$, with the maximally entangled state vectors corresponding to $\kappa_1,...,\kappa_4$, being the eigenvectors.

Interpreting the cost as the time taken for an elementary transition to take place, we can use $k_1=0$ (no time is taken to do nothing) and $k_4=k_2+k_3$ (the total time for $U_2$ followed by $U_3$), with $k_2,k_3>0$. For convenience, we scale this to
\[
k_2=k
\text{ and }
k_3=2,
\]
with $C$ then having eigenvalues $0$, $k$, $2$ and $k+2$.

Without any constraints we indeed find that the optimal channel is given by $U_1$, with cost $0$. 

Let us require the channel to satisfy
\[
\mathcal{E}
\left[ 
\begin{array}{cc}
1 & 0 \\ 
0 & 0 
\end{array}
\right]
=
\left[ 
\begin{array}{cc}
0 & 0 \\ 
0 & 1 
\end{array}
\right].
\]

For $k\leq1$ the optimal channel $\mathcal{E}$ is given by $U_2$, with cost
\[
K_C(\mathcal{E})=k,
\] 
corresponding to the maximally entangled eigenstate $\kappa_2$ of $C$. The fact that it is maximally entangled, thus representing a channel by duality, is exactly what allows the corresponding cost to be reached. The same is true for $U_1$ in the unconstrained case above. This is compatible with  Section \ref{AfdRolVanVerstr}'s expectation that entanglement in the lower cost eigenvectors of $C$, tends to lower the optimal cost.

However, $U_4$ also satisfies the constraint, so $U_3$ may enter the mix if $U_2$'s cost becomes high enough. Indeed, for $k\geq 1$, the optimal channel is
\begin{equation}
\mathcal{E}
\left[ 
\begin{array}{cc}
\rho_{11} & \rho_{12} \\ 
\rho_{21} & \rho_{22} 
\end{array}
\right]
=
\left[ 
\begin{array}{cc}
\frac{1}{k^2}\rho_{22} & \frac{1}{k}\rho_{21} \\ 
\frac{1}{k}\rho_{12}   & \rho_{11}+\left(1-\frac{1}{k^2}\right)\rho_{22} 
\end{array}
\right]
\label{tydKanaal}
\end{equation}
with cost 
\[
K_C(\mathcal{E})=1+\frac{k}{2}-\frac{1}{2k}.
\]
Note that for $k\neq 1$ this cost is indeed lower than $U_2$'s cost $k$. But, for $k<1$, the map $\mathcal{E}$ in (\ref{tydKanaal}) is no longer a channel, explaining the need for another map, namely the channel given by $U_2$.

Decomposing the optimal channel (\ref{tydKanaal}) into elementary transitions, one obtains 
\begin{equation}
\mathcal{E}=
\frac{1}{2}\left(1-\frac{1}{k^2}\right)\mathcal{E}_1+
\frac{1}{2}\left(1+\frac{1}{k^2}\right)\mathcal{E}_2,
\label{ontbindVb}
\end{equation}
where $\mathcal{E}_1$, given by
\[
\mathcal{E}_1
\left[ 
\begin{array}{cc}
\rho_{11} & \rho_{12} \\ 
\rho_{21} & \rho_{22} 
\end{array}
\right]
=
\left[ 
\begin{array}{cc}
0 & 0 \\ 
0 & 2\rho_{22} 
\end{array}
\right],
\]
is dual to the separable vector state $\ket{2}\ket{2}$, and $\mathcal{E}_2$, given by
\[
\mathcal{E}_2
\left[ 
\begin{array}{cc}
\rho_{11} & \rho_{12} \\ 
\rho_{21} & \rho_{22} 
\end{array}
\right]
=
\frac{2}{1+k^2}
\left[ 
\begin{array}{cc}
\rho_{22}  & k\rho_{21} \\ 
k\rho_{12} & k^2\rho_{11}
\end{array}
\right],
\]
is dual to the entangled vector state $\ket{1}\ket{2}+\frac{1}{k}\ket{2}\ket{1}$, not displayed normalized here.

The cost of $\mathcal{E}_1$ is $1$, whereas $\mathcal{E}_2$ costs $(k+1-1/k+1/k^2)/(1+1/k^2)>1$. 
So $\mathcal{E}_1$, while having no entanglement, lowers the cost in the weighted average $K_C(\mathcal{E})$ of these two costs given by the probability weights appearing in (\ref{ontbindVb}). 

On the other hand, the more expensive elementary transition $\mathcal{E}_2$, being dual to an entangled state, indeed carries the larger weight, as expected in Section \ref{AfdRolVanVerstr}. This state becomes maximally entangled for $k=1$, in which case $\mathcal{E}_2$ is given by $U_2$ and  the weight of $\mathcal{E}_1$ in $\mathcal{E}$ becomes zero, in line with the case $k\leq1$ above. 

As for the general nature of the elementary transitions $\mathcal{E}_1$ and $\mathcal{E}_2$: For $k>1$, the optimal channel is not unitary, and neither are $\mathcal{E}_1$ or $\mathcal{E}_2$. The support of $\mathcal{E}_1$ is spanned by $\ket{2}$. Restricting $\mathcal{E}_1$ to this space and dividing by 2, gives a trivial channel. The support of $\mathcal{E}_2$ is $H_A$, but no scalar multiple makes it a channel, unless $k=1$, in which case $\mathcal{E}_2$ itself is a unitary channel given by $U_2$. 

In the limit where $k\rightarrow\infty$, both elementary transitions $\mathcal{E}_1$ and $\mathcal{E}_2$ are dual to separable states, namely $\ket{2}\ket{2}$ and $\ket{1}\ket{2}$ respectively. 
In this case both have relatively large weights of $1/2$ each, indicating that our heuristic arguments in Section \ref{AfdRolVanVerstr} are not quite the full story. A more detailed study of how large the weight of separable state can be in a channel decomposition, would be required to fill the holes. We do not pursue that here.

\section{Minimal disturbance}\label{AfdMinVer}

Here our goal is to outline a method to theoretically design or search for a channel from a system $A$ to itself, which performs a specified set of tasks,
\begin{equation}
\mathcal{E}(\rho^\text{in}_j)=\rho^\text{out}_j\text{ for }j=1,...,l,
\label{take}
\end{equation}
 while otherwise disturbing the state of $A$ as little as possible.

We formulate this more precisely, by casting it as an optimal channel problem: 

Consider any set
\[
g_1,...,g_v
\] 
of self-adjoint $m\times m$ matrices which generate $M_m$ as an algebra, again in terms of the orthonormal basis $\ket{1},...,\ket{m}$ for the Hilbert space $H_A$. Set up the cost matrix
\[
C=
\sum_{j=1}^{v}|I_m\otimes g_j-g_j^T\otimes I_m|^2
\] 
in terms of them. (Here $|O|^2=O^\dagger O$.) One then  searches for an optimal channel satisfying (\ref{take}), in terms of this cost matrix.

\subsection{Motivating $C$}

This cost is an analogue of distance squared in $\mathbb{R}^{v}$, with the generators in place of the coordinate functions. However, we do not require the set of generators to be a minimal set generating $M_m$. We allow for the possibility that some strict subset of the generators $g_1,...,g_v$, generate the whole of $M_m$. The relevance of different choices of generators will be discussed in the next subsection.

Note that in absence of (\ref{take}), the optimal cost is zero. To see this, note that
\[
(I_m\otimes g-g^T\otimes I_m)\ket{\Omega}=0
\]
for any $m\times m$ matrix $g$, as is easily checked from the definition (\ref{Omega}) of the maximally entangled state vector $\ket{\Omega}$. 
Consequently, $\ket{\Omega}$ is an eigenvector of $C$ with eigenvalue $0$, which is necessarily the smallest eigenvalue of $C$, as the latter is a positive operator. 

In particular, for the identity channel $\mathcal{E}$, for which $\kappa_\mathcal{E}=\ket{\Omega}\bra{\Omega}$, we have
\[
K_C(\mathcal{E})=\bra{\Omega}C\ket{\Omega}=0.
\]

Conversely, using the theory of the quadratic Wasserstein metric associated to the generators, one finds that the identity channel 
is the unique optimal channel. 
This is given by Corollary 6.4 of \cite{D}, as the unique zero cost transport plan from the maximally mixed state $\frac{1}{m}\Tr_m$ of $A$, to itself. Recall from Section \ref{AfdElemOorg} that $\frac{1}{m}\Tr_m$ is our reference state.
(In \cite{D}, the transposition in $C$ appears in a more general form via an operator $S$, associated in this case to $\frac{1}{m}\Tr_m$ and specializing to the transposition.)

This shows that the cost matrix $C$ above, tends to force the channel towards the identity channel, i.e., towards causing minimal disturbance in the state of $A$.

\subsection{The significance of the generator sets}

Different choices of generator sets can in effect weigh the cost of elementary transitions differently. We illustrate this for a system $A$ consisting of $r$ spins, giving $m=2^r$.

Take the orthonormal basis used in the Choi-Jamio{\l}kowski duality as
\begin{align*}
\ket{0}  & =\ket{-}...\ket{-},\\
\ket{1}  & =\ket{-}...\ket{-}\ket{+},\\
\ket{2}  & =\ket{-}...\ket{-}\ket{+}\ket{-},\\
& \vdots\\
\ket{m-1}  & =\ket{+}...\ket{+},
\end{align*}
with $\ket{\pm}$ the up/down $z$-spin states.

One set of generators for $M_m$ in this case, is given by the spin observables
\[
g_{ij}=
I_2^{(1)}\otimes...\otimes I_2^{(i-1)}
\otimes\sigma_j^{(i)}
\otimes I_2^{(i+1)}...\otimes I_2^{(r)},
\]
with $\sigma_j$ in the $i$'th position, for the $x$, $y$ and $z$ Pauli matrices $\sigma_1$, $\sigma_2$ and $\sigma_3$, where $i=1,...,r$ and $j=1,2,3$. (The superscripts in brackets simply indicate the position in the elementary tensor.)

A second set, $h_1,h_2$, is given by
\[
h_1=
\left[
\begin{array}
[c]{ccccc}
0 &  &  &  & \\
& 1 &  &  & \\
&  & 2 &  & \\
&  &  & \ddots & \\
&  &  &  & m-1
\end{array}
\right],
\]
(expressed in the basis $\ket{0},...,\ket{m-1}$) together with its Fourier transform 
\[
h_2=F^{\dagger}h_1 F
\] 
where $F$ is the Fourier
transform on $\mathbb{Z}_{m}$, also called the quantum Fourier transform, namely the $m\times m$ 
unitary matrix $F_{jk}=e^{-2\pi jk/m}/m^{1/2}$ for $j,k=0,1,...,m-1$.
Heuristically, we can think of $h_1$ and $h_2$ as discrete position and momentum respectively, since they are analogous to position and momentum in one dimension in the usual continuous case (also see the old paper by Schwinger \cite{S} for the unitary representation of this, as well as \cite{RMG,MRG,MSG}). This analogy should make physical sense in the large $m$ limit.

Changing one spin value in the basis, can change the state from $\ket{0}$ to $\ket{1}$, but it can also change the state from $\ket{0}$ to $\ket{2^{r-1}}$, for example. 
The difference in the jumps in the state labels, is simply an artefact of our choice of representation of the basis states.

When using the cost matrix
\[
C_h=
|I_m\otimes h_1-h_1\otimes I_m|^2
+
|I_m\otimes h_2-h_2^T\otimes I_m|^2,
\]  
we can correspondingly expect that the cost is not balanced between the spins, with changes at the left of our representation of the spins in the basis states, being suppressed compared to those at the right, due to higher cost coming from the term $|I_m\otimes h_1-h_1\otimes I_m|^2$.

To spread the cost evenly between the spins, making it independent of our specific mathematical representation in the basis states, and physically more sensible, one should rather use the cost matrix
\[
C_g=
\sum_{ij}|I_m\otimes g_{ij}-g_{ij}^T\otimes I_m|^2.
\] 

The cost matrix $C_h$ would be more suitable if we interpret the states $\ket{0},...,\ket{m-1}$ as being physically further removed in some sense, when the difference in their labels is large. For example, if the labels $0,...,m-1$, being the values of the observable $h_1$, form a discrete set of positions in a straight line, while the corresponding labels for the second observable, $h_2$, is thought of as a discrete set of momenta. The cost $C_h$ then reflects the size of the difference between the labels.

\section{Concluding remarks} \label{AfdAfsl}
We have presented a framework to optimize quantum channels. It uses the Choi-Jamio{\l}kowski duality to decompose a channel into elementary transitions, which are analogous to point to point transitions in classical optimal transport. This is a conceptually satisfying setup for optimization.

Our main conclusion regarding the general setup, is that entanglement in the lower lying cost eigenvectors, aids in lowering the cost, i.e., tends to allow for cheaper channels. We saw this through mostly heuristic arguments in Section \ref{AfdRolVanVerstr}, as well as by example in Section \ref{AfdVbe}. 
This point was also illustrated in Section \ref{AfdMinVer}, for a specific class of cost matrices, in arbitrary dimensions. 
The second example in Section \ref{AfdVbe}, however, also showed the quantitative limitations of the general heuristic arguments.

As an application, we considered how one can obtain channels that perform a prescribed set of tasks, while otherwise disturbing the state as little as possible in terms if the chosen cost matrix. We expect that this should be of value in quantum information processing.

In this application, we used self-adjoint operators as generators, but this can be generalized at least to the case where the set of generators $g_1,...,g_v$ collectively is self-adjoint, meaning that 
$\lbrace g_1^\dagger,...,g_v^\dagger\rbrace=\lbrace g_1,...,g_v\rbrace$. 
Then the identity channel is still the unique optimal channel, as can be seen in \cite{D}.

Technical work that remains, includes making the heuristic arguments in Section \ref{AfdRolVanVerstr} more precise and quantitative. 
One could also explore more quantitatively how much the states, other than the specified input states in (\ref{take}), are changed by optimal channels obtained in Section \ref{AfdMinVer}'s setup, including how this depends on the distance of the state in question from those input states, and on the set of generators.

If in Section \ref{AfdMinVer} one can check from $C$'s definition that $\ket{\Omega}$ is its only eigenvector (up to scalar multiple) with eigenvalue $0$, then it would provide a second method, independent of the Wasserstein metric, to show that the identity channel is the unique optimal channel in absence of (\ref{take}). This may provide a way to generalize the allowed generator sets beyond the condition
$\lbrace g_1^\dagger,...,g_v^\dagger\rbrace=\lbrace g_1,...,g_v\rbrace$ mentioned above,
while still ensuring a unique optimal channel.

Furthermore, investigating computational techniques to determine the optimal cost and channels, for example in Section \ref{AfdMinVer}, would also be of much value. As the dimension of the Hilbert spaces increase, this can be expected to become challenging. 

Further examples, or classes, of cost matrices, should be explored, as this paper only looked at a limited selection, for two copies of the same system.

A more general aspect of our approach that appears worth developing further, is the refined picture of an elementary transition in Section \ref{AfdElemOorg}, in terms of its support. This may give deeper insight into the structure of a channel via a decomposition into elementary transitions, in particular for an optimal channel.

\begin{acknowledgments}
I thank the referees for a number of valuable suggestions which improved the overall structure and presentation of the paper.
\end{acknowledgments}


%

{\tiny }
\begin{thebibliography}{36}%
	\makeatletter
	\providecommand \@ifxundefined [1]{%
		\@ifx{#1\undefined}
	}%
	\providecommand \@ifnum [1]{%
		\ifnum #1\expandafter \@firstoftwo
		\else \expandafter \@secondoftwo
		\fi
	}%
	\providecommand \@ifx [1]{%
		\ifx #1\expandafter \@firstoftwo
		\else \expandafter \@secondoftwo
		\fi
	}%
	\providecommand \natexlab [1]{#1}%
	\providecommand \enquote  [1]{``#1''}%
	\providecommand \bibnamefont  [1]{#1}%
	\providecommand \bibfnamefont [1]{#1}%
	\providecommand \citenamefont [1]{#1}%
	\providecommand \href@noop [0]{\@secondoftwo}%
	\providecommand \href [0]{\begingroup \@sanitize@url \@href}%
	\providecommand \@href[1]{\@@startlink{#1}\@@href}%
	\providecommand \@@href[1]{\endgroup#1\@@endlink}%
	\providecommand \@sanitize@url [0]{\catcode `\\12\catcode `\$12\catcode
		`\&12\catcode `\#12\catcode `\^12\catcode `\_12\catcode `\%12\relax}%
	\providecommand \@@startlink[1]{}%
	\providecommand \@@endlink[0]{}%
	\providecommand \url  [0]{\begingroup\@sanitize@url \@url }%
	\providecommand \@url [1]{\endgroup\@href {#1}{\urlprefix }}%
	\providecommand \urlprefix  [0]{URL }%
	\providecommand \Eprint [0]{\href }%
	\providecommand \doibase [0]{https://doi.org/}%
	\providecommand \selectlanguage [0]{\@gobble}%
	\providecommand \bibinfo  [0]{\@secondoftwo}%
	\providecommand \bibfield  [0]{\@secondoftwo}%
	\providecommand \translation [1]{[#1]}%
	\providecommand \BibitemOpen [0]{}%
	\providecommand \bibitemStop [0]{}%
	\providecommand \bibitemNoStop [0]{.\EOS\space}%
	\providecommand \EOS [0]{\spacefactor3000\relax}%
	\providecommand \BibitemShut  [1]{\csname bibitem#1\endcsname}%
	\let\auto@bib@innerbib\@empty
\bibitem [{\citenamefont {Csisz{\'a}r}\ \emph {et~al.}(2007)\citenamefont
	{Csisz{\'a}r}, \citenamefont {Hiai},\ and\ \citenamefont {Petz}}]{CHP}%
\BibitemOpen
\bibfield  {author} {\bibinfo {author} {\bibfnamefont {I.}~\bibnamefont
		{Csisz{\'a}r}}, \bibinfo {author} {\bibfnamefont {F.}~\bibnamefont {Hiai}},\
	and\ \bibinfo {author} {\bibfnamefont {D.}~\bibnamefont {Petz}},\ }\href@noop
{} {\bibfield  {journal} {\bibinfo  {journal} {J. Math. Phys.}\ }\textbf
	{\bibinfo {volume} {48}},\ \bibinfo {pages} {092102} (\bibinfo {year}
	{2007})}\BibitemShut {NoStop}%
\bibitem [{\citenamefont {Jarzyna}(2017)}]{Jar}%
\BibitemOpen
\bibfield  {author} {\bibinfo {author} {\bibfnamefont {M.}~\bibnamefont
		{Jarzyna}},\ }\href@noop {} {\bibfield  {journal} {\bibinfo  {journal} {Phys.
			Rev. A}\ }\textbf {\bibinfo {volume} {96}},\ \bibinfo {pages} {032340}
	(\bibinfo {year} {2017})}\BibitemShut {NoStop}%
\bibitem [{\citenamefont {Ding}\ \emph {et~al.}(2019)\citenamefont {Ding},
	\citenamefont {Pavlichin},\ and\ \citenamefont {Wilde}}]{DPW}%
\BibitemOpen
\bibfield  {author} {\bibinfo {author} {\bibfnamefont {D.}~\bibnamefont
		{Ding}}, \bibinfo {author} {\bibfnamefont {D.~S.}\ \bibnamefont
		{Pavlichin}},\ and\ \bibinfo {author} {\bibfnamefont {M.~M.}\ \bibnamefont
		{Wilde}},\ }\href@noop {} {\bibfield  {journal} {\bibinfo  {journal} {IEEE
			Trans. Inf. Theory}\ }\textbf {\bibinfo {volume} {65}},\ \bibinfo {pages}
	{418} (\bibinfo {year} {2019})}\BibitemShut {NoStop}%
\bibitem [{\citenamefont {de~Palma}\ and\ \citenamefont {Trevisan}()}]{DePT}%
\BibitemOpen
\bibfield  {author} {\bibinfo {author} {\bibfnamefont {G.}~\bibnamefont
		{de~Palma}}\ and\ \bibinfo {author} {\bibfnamefont {D.}~\bibnamefont
		{Trevisan}},\ }\href@noop {} \emph{Quantum Optimal Transport with Quantum
	Channels}, {\bibfield  {journal} {\bibinfo  {journal} {Ann.
			Henri Poincar{\'e}}\ }}\bibinfo {note} {(to appear)},\ \Eprint
{https://arxiv.org/abs/arXiv:1911.00803} {arXiv:1911.00803v2} \BibitemShut
{NoStop}%
\bibitem [{\citenamefont {de~Pillis}(1967)}]{deP}%
\BibitemOpen
\bibfield  {author} {\bibinfo {author} {\bibfnamefont {J.}~\bibnamefont
		{de~Pillis}},\ }\href@noop {} {\bibfield  {journal} {\bibinfo  {journal}
		{Pac. J. Math.}\ }\textbf {\bibinfo {volume} {23}},\ \bibinfo {pages} {129}
	(\bibinfo {year} {1967})}\BibitemShut {NoStop}%
\bibitem [{\citenamefont {Jamio{\l}kowski}(1972)}]{J}%
\BibitemOpen
\bibfield  {author} {\bibinfo {author} {\bibfnamefont {A.}~\bibnamefont
		{Jamio{\l}kowski}},\ }\href@noop {} {\bibfield  {journal} {\bibinfo
		{journal} {Rep. Math. Phys.}\ }\textbf {\bibinfo {volume} {3}},\ \bibinfo
	{pages} {275} (\bibinfo {year} {1972})}\BibitemShut {NoStop}%
\bibitem [{\citenamefont {Choi}(1975)}]{C}%
\BibitemOpen
\bibfield  {author} {\bibinfo {author} {\bibfnamefont {M.-D.}\ \bibnamefont
		{Choi}},\ }\href@noop {} {\bibfield  {journal} {\bibinfo  {journal} {Linear
			Alg. Appl.}\ }\textbf {\bibinfo {volume} {10}},\ \bibinfo {pages} {285}
	(\bibinfo {year} {1975})}\BibitemShut {NoStop}%
\bibitem [{\citenamefont {Duvenhage}()}]{D}%
\BibitemOpen
\bibfield  {author} {\bibinfo {author} {\bibfnamefont {R.}~\bibnamefont
		{Duvenhage}},\ }\href@noop {} \emph{Quadratic Wasserstein metrics for von Neumann algebras via transport plans}, {\bibfield  {journal} {\bibinfo  {journal} {J.
			Operator Theory}\ }}\bibinfo {note} {(to appear)},\ \Eprint
{https://arxiv.org/abs/arXiv:2012.03564v3} {arXiv:2012.03564v3} \BibitemShut
{NoStop}%
\bibitem [{\citenamefont {{\.Z}yczkowski}\ and\ \citenamefont
	{S{\l}omczy{\'n}ski}(1998)}]{ZS}%
\BibitemOpen
\bibfield  {author} {\bibinfo {author} {\bibfnamefont {K.}~\bibnamefont
		{{\.Z}yczkowski}}\ and\ \bibinfo {author} {\bibfnamefont {W.}~\bibnamefont
		{S{\l}omczy{\'n}ski}},\ }\href@noop {} {\bibfield  {journal} {\bibinfo
		{journal} {J. Phys. A}\ }\textbf {\bibinfo {volume} {31}},\ \bibinfo {pages}
	{9095} (\bibinfo {year} {1998})}\BibitemShut {NoStop}%
\bibitem [{\citenamefont {Carlen}\ and\ \citenamefont {Maas}(2014)}]{CM1}%
\BibitemOpen
\bibfield  {author} {\bibinfo {author} {\bibfnamefont {E.~A.}\ \bibnamefont
		{Carlen}}\ and\ \bibinfo {author} {\bibfnamefont {J.}~\bibnamefont {Maas}},\
}\href@noop {} {\bibfield  {journal} {\bibinfo  {journal} {Comm. Math.
			Phys.}\ }\textbf {\bibinfo {volume} {331}},\ \bibinfo {pages} {887} (\bibinfo
	{year} {2014})}\BibitemShut {NoStop}%
\bibitem [{\citenamefont {Carlen}\ and\ \citenamefont {Maas}(2017)}]{CM2}%
\BibitemOpen
\bibfield  {author} {\bibinfo {author} {\bibfnamefont {E.~A.}\ \bibnamefont
		{Carlen}}\ and\ \bibinfo {author} {\bibfnamefont {J.}~\bibnamefont {Maas}},\
}\href@noop {} {\bibfield  {journal} {\bibinfo  {journal} {J. Funct. Anal.}\
	}\textbf {\bibinfo {volume} {273}},\ \bibinfo {pages} {1810} (\bibinfo {year}
	{2017})}\BibitemShut {NoStop}%
\bibitem [{\citenamefont {Carlen}\ and\ \citenamefont {Maas}(2020)}]{CM3}%
\BibitemOpen
\bibfield  {author} {\bibinfo {author} {\bibfnamefont {E.~A.}\ \bibnamefont
		{Carlen}}\ and\ \bibinfo {author} {\bibfnamefont {J.}~\bibnamefont {Maas}},\
}\href@noop {} {\bibfield  {journal} {\bibinfo  {journal} {J. Stat. Phys.}\
	}\textbf {\bibinfo {volume} {178}},\ \bibinfo {pages} {319} (\bibinfo {year}
	{2020})}\BibitemShut {NoStop}%
\bibitem [{\citenamefont {Yamamoto}\ \emph {et~al.}(2018)\citenamefont
	{Yamamoto}, \citenamefont {Yongxin}, \citenamefont {Georgiou},\ and\
	\citenamefont {Tannenbaum}}]{YYGT}%
\BibitemOpen
\bibfield  {author} {\bibinfo {author} {\bibfnamefont {K.}~\bibnamefont
		{Yamamoto}}, \bibinfo {author} {\bibfnamefont {N.}~\bibnamefont {Yongxin}},
	\bibinfo {author} {\bibfnamefont {T.~T.}\ \bibnamefont {Georgiou}},\ and\
	\bibinfo {author} {\bibfnamefont {A.}~\bibnamefont {Tannenbaum}},\
}\href@noop {} {\bibfield  {journal} {\bibinfo  {journal} {IEEE Trans.
			Automat. Control}\ }\textbf {\bibinfo {volume} {63}},\ \bibinfo {pages}
	{1208} (\bibinfo {year} {2018})}\BibitemShut {NoStop}%
\bibitem [{\citenamefont {Chen}\ \emph {et~al.}(2018)\citenamefont {Chen},
	\citenamefont {Georgiou},\ and\ \citenamefont {Tannenbaum}}]{CGT}%
\BibitemOpen
\bibfield  {author} {\bibinfo {author} {\bibfnamefont {Y.}~\bibnamefont
		{Chen}}, \bibinfo {author} {\bibfnamefont {T.~T.}\ \bibnamefont {Georgiou}},\
	and\ \bibinfo {author} {\bibfnamefont {A.}~\bibnamefont {Tannenbaum}},\
}\href@noop {} {\bibfield  {journal} {\bibinfo  {journal} {IEEE Trans.
			Automat. Control}\ }\textbf {\bibinfo {volume} {63}},\ \bibinfo {pages}
	{2612} (\bibinfo {year} {2018})}\BibitemShut {NoStop}%
\bibitem [{\citenamefont {Chen}\ \emph {et~al.}(2020)\citenamefont {Chen},
	\citenamefont {Gangbo}, \citenamefont {Georgiou},\ and\ \citenamefont
	{Tannenbaum}}]{CGGT}%
\BibitemOpen
\bibfield  {author} {\bibinfo {author} {\bibfnamefont {Y.}~\bibnamefont
		{Chen}}, \bibinfo {author} {\bibfnamefont {W.}~\bibnamefont {Gangbo}},
	\bibinfo {author} {\bibfnamefont {T.~T.}\ \bibnamefont {Georgiou}},\ and\
	\bibinfo {author} {\bibfnamefont {A.}~\bibnamefont {Tannenbaum}},\
}\href@noop {} {\bibfield  {journal} {\bibinfo  {journal} {European J. Appl.
			Math.}\ }\textbf {\bibinfo {volume} {31}},\ \bibinfo {pages} {574} (\bibinfo
	{year} {2020})}\BibitemShut {NoStop}%
\bibitem [{\citenamefont {Golse}\ \emph {et~al.}(2016)\citenamefont {Golse},
	\citenamefont {Mouhot},\ and\ \citenamefont {Paul}}]{GMP}%
\BibitemOpen
\bibfield  {author} {\bibinfo {author} {\bibfnamefont {F.}~\bibnamefont
		{Golse}}, \bibinfo {author} {\bibfnamefont {C.}~\bibnamefont {Mouhot}},\ and\
	\bibinfo {author} {\bibfnamefont {T.}~\bibnamefont {Paul}},\ }\href@noop {}
{\bibfield  {journal} {\bibinfo  {journal} {Comm. Math. Phys.}\ }\textbf
	{\bibinfo {volume} {343}},\ \bibinfo {pages} {165} (\bibinfo {year}
	{2016})}\BibitemShut {NoStop}%
\bibitem [{\citenamefont {Golse}\ and\ \citenamefont {Paul}(2018)}]{GP}%
\BibitemOpen
\bibfield  {author} {\bibinfo {author} {\bibfnamefont {F.}~\bibnamefont
		{Golse}}\ and\ \bibinfo {author} {\bibfnamefont {T.}~\bibnamefont {Paul}},\
}\href@noop {} {\bibfield  {journal} {\bibinfo  {journal} {C. R. Math. Acad.
			Sci. Paris}\ }\textbf {\bibinfo {volume} {356}},\ \bibinfo {pages} {177}
	(\bibinfo {year} {2018})}\BibitemShut {NoStop}%
\bibitem [{\citenamefont {Caglioti}\ \emph {et~al.}(2020)\citenamefont
	{Caglioti}, \citenamefont {Golse},\ and\ \citenamefont {Paul}}]{CGP1}%
\BibitemOpen
\bibfield  {author} {\bibinfo {author} {\bibfnamefont {E.}~\bibnamefont
		{Caglioti}}, \bibinfo {author} {\bibfnamefont {F.}~\bibnamefont {Golse}},\
	and\ \bibinfo {author} {\bibfnamefont {T.}~\bibnamefont {Paul}},\ }\href@noop
{} {\bibfield  {journal} {\bibinfo  {journal} {J. Stat. Phys.}\ }\textbf
	{\bibinfo {volume} {181}},\ \bibinfo {pages} {149} (\bibinfo {year}
	{2020})}\BibitemShut {NoStop}%
\bibitem [{\citenamefont {Caglioti}\ \emph {et~al.}(2021)\citenamefont
	{Caglioti}, \citenamefont {Golse},\ and\ \citenamefont {Paul}}]{CGP2}%
\BibitemOpen
\bibfield  {author} {\bibinfo {author} {\bibfnamefont {E.}~\bibnamefont
		{Caglioti}}, \bibinfo {author} {\bibfnamefont {F.}~\bibnamefont {Golse}},\
	and\ \bibinfo {author} {\bibfnamefont {T.}~\bibnamefont {Paul}},\ }\href@noop
{} \emph{Towards optimal transport for quantum densities}, {} (\bibinfo {year} {2021}),\ \Eprint
{https://arxiv.org/abs/arXiv:2101.03256v2} {arXiv:2101.03256v2} \BibitemShut
{NoStop}%
\bibitem [{\citenamefont {Agredo}\ and\ \citenamefont {Fagnola}(2017)}]{AF}%
\BibitemOpen
\bibfield  {author} {\bibinfo {author} {\bibfnamefont {J.}~\bibnamefont
		{Agredo}}\ and\ \bibinfo {author} {\bibfnamefont {F.}~\bibnamefont
		{Fagnola}},\ }\href@noop {} {\bibfield  {journal} {\bibinfo  {journal}
		{Stochastics}\ }\textbf {\bibinfo {volume} {89}},\ \bibinfo {pages} {910}
	(\bibinfo {year} {2017})}\BibitemShut {NoStop}%
\bibitem [{\citenamefont {Peyr{\'e}}\ \emph {et~al.}(2019)\citenamefont
	{Peyr{\'e}}, \citenamefont {Chizat}, \citenamefont {Vialard},\ and\
	\citenamefont {Solomon}}]{PCVS}%
\BibitemOpen
\bibfield  {author} {\bibinfo {author} {\bibfnamefont {G.}~\bibnamefont
		{Peyr{\'e}}}, \bibinfo {author} {\bibfnamefont {L.}~\bibnamefont {Chizat}},
	\bibinfo {author} {\bibfnamefont {F.-X.}\ \bibnamefont {Vialard}},\ and\
	\bibinfo {author} {\bibfnamefont {J.}~\bibnamefont {Solomon}},\ }\href@noop
{} {\bibfield  {journal} {\bibinfo  {journal} {European J. Appl. Math.}\
	}\textbf {\bibinfo {volume} {30}},\ \bibinfo {pages} {1079} (\bibinfo {year}
	{2019})}\BibitemShut {NoStop}%
\bibitem [{\citenamefont {Datta}\ and\ \citenamefont {Rouz{\'e}}(2020)}]{DR}%
\BibitemOpen
\bibfield  {author} {\bibinfo {author} {\bibfnamefont {N.}~\bibnamefont
		{Datta}}\ and\ \bibinfo {author} {\bibfnamefont {C.}~\bibnamefont
		{Rouz{\'e}}},\ }\href@noop {} {\bibfield  {journal} {\bibinfo  {journal}
		{Ann. Henri Poincar{\'e}}\ }\textbf {\bibinfo {volume} {21}},\ \bibinfo
	{pages} {2115} (\bibinfo {year} {2020})}\BibitemShut {NoStop}%
\bibitem [{\citenamefont {Ikeda}(2020)}]{Ik}%
\BibitemOpen
\bibfield  {author} {\bibinfo {author} {\bibfnamefont {K.}~\bibnamefont
		{Ikeda}},\ }\href@noop {} {\bibfield  {journal} {\bibinfo  {journal} {Quantum
			Inf. Process.}\ }\textbf {\bibinfo {volume} {19}},\ \bibinfo {pages} {25}
	(\bibinfo {year} {2020})}\BibitemShut {NoStop}%
\bibitem [{\citenamefont {de~Palma}\ \emph {et~al.}()\citenamefont {de~Palma},
	\citenamefont {Marvian}, \citenamefont {Trevisan},\ and\ \citenamefont
	{Lloyd}}]{dePMTL}%
\BibitemOpen
\bibfield  {author} {\bibinfo {author} {\bibfnamefont {G.}~\bibnamefont
		{de~Palma}}, \bibinfo {author} {\bibfnamefont {M.}~\bibnamefont {Marvian}},
	\bibinfo {author} {\bibfnamefont {D.}~\bibnamefont {Trevisan}},\ and\
	\bibinfo {author} {\bibfnamefont {S.}~\bibnamefont {Lloyd}},\ }\href@noop {} \emph{The quantum Wasserstein distance of order 1}, 
{\bibfield  {journal} {\bibinfo  {journal} {IEEE Trans. Inf. Theory}\
}}\bibinfo {note} {(to appear)},\ \Eprint
{https://arxiv.org/abs/arXiv:2009.04469v1} {arXiv:2009.04469v2} \BibitemShut
{NoStop}%
\bibitem [{\citenamefont {Kiani}\ \emph {et~al.}(2021)\citenamefont {Kiani},
	\citenamefont {de~Palma}, \citenamefont {Marvian}, \citenamefont {Liu},\ and\
	\citenamefont {Lloyd}}]{KdePMLL}%
\BibitemOpen
\bibfield  {author} {\bibinfo {author} {\bibfnamefont {B.~T.}\ \bibnamefont
		{Kiani}}, \bibinfo {author} {\bibfnamefont {G.}~\bibnamefont {de~Palma}},
	\bibinfo {author} {\bibfnamefont {M.}~\bibnamefont {Marvian}}, \bibinfo
	{author} {\bibfnamefont {Z.-W.}\ \bibnamefont {Liu}},\ and\ \bibinfo {author}
	{\bibfnamefont {S.}~\bibnamefont {Lloyd}},\ }\href@noop {} \emph{Quantum Earth Mover's Distance: A New Approach to
Learning Quantum Data}, {} (\bibinfo
{year} {2021}),\ \Eprint {https://arxiv.org/abs/arXiv:2101.03037v1}
{arXiv:2101.03037v1} \BibitemShut {NoStop}%
\bibitem [{\citenamefont {Friedland}\ \emph {et~al.}(2021)\citenamefont
	{Friedland}, \citenamefont {Eckstein}, \citenamefont {Cole},\ and\
	\citenamefont {{\.Z}yczkowski}}]{FECZ}%
\BibitemOpen
\bibfield  {author} {\bibinfo {author} {\bibfnamefont {S.}~\bibnamefont
		{Friedland}}, \bibinfo {author} {\bibfnamefont {M.}~\bibnamefont {Eckstein}},
	\bibinfo {author} {\bibfnamefont {S.}~\bibnamefont {Cole}},\ and\ \bibinfo
	{author} {\bibfnamefont {K.}~\bibnamefont {{\.Z}yczkowski}},\ }\href@noop {} \emph{Quantum Monge-Kantorovich problem and transport distance between density matrices},
{\  (\bibinfo {year} {2021})},\ \Eprint
{https://arxiv.org/abs/arXiv:2102.07787v1} {arXiv:2102.07787v1} \BibitemShut
{NoStop}%
\bibitem [{\citenamefont {Cole}\ \emph {et~al.}(2021)\citenamefont {Cole},
	\citenamefont {Eckstein}, \citenamefont {Friedland},\ and\ \citenamefont
	{{\.Z}yczkowski}}]{CEFZ}%
\BibitemOpen
\bibfield  {author} {\bibinfo {author} {\bibfnamefont {S.}~\bibnamefont
		{Cole}}, \bibinfo {author} {\bibfnamefont {M.}~\bibnamefont {Eckstein}},
	\bibinfo {author} {\bibfnamefont {S.}~\bibnamefont {Friedland}},\ and\
	\bibinfo {author} {\bibfnamefont {K.}~\bibnamefont {{\.Z}yczkowski}},\
}\href@noop {} \emph{Quantum optimal transport}, {} (\bibinfo {year} {2021}),\ \Eprint
{https://arxiv.org/abs/arXiv:2105.06922v1} {arXiv:2105.06922v1} \BibitemShut
{NoStop}%
\bibitem [{\citenamefont {Villani}(2003)}]{V}%
\BibitemOpen
\bibfield  {author} {\bibinfo {author} {\bibfnamefont {C.}~\bibnamefont
		{Villani}},\ }\href@noop {} {\emph {\bibinfo {title} {Topics in optimal
			transportation}}}\ (\bibinfo  {publisher} {American Mathematical Society,
	Providence},\ \bibinfo {year} {2003})\BibitemShut {NoStop}%
\bibitem [{\citenamefont {Datta}(2005)}]{Dat}%
\BibitemOpen
\bibfield  {author} {\bibinfo {author} {\bibfnamefont {S.}~\bibnamefont
		{Datta}},\ }\href@noop {} {\emph {\bibinfo {title} {Quantum Transport: Atom
			to Transistor}}}\ (\bibinfo  {publisher} {Cambridge University Press},\
\bibinfo {year} {2005})\BibitemShut {NoStop}%
\bibitem [{\citenamefont {Jiang}\ \emph {et~al.}(2013)\citenamefont {Jiang},
	\citenamefont {Luo},\ and\ \citenamefont {Fu}}]{JLF}%
\BibitemOpen
\bibfield  {author} {\bibinfo {author} {\bibfnamefont {M.}~\bibnamefont
		{Jiang}}, \bibinfo {author} {\bibfnamefont {S.}~\bibnamefont {Luo}},\ and\
	\bibinfo {author} {\bibfnamefont {S.}~\bibnamefont {Fu}},\ }\href@noop {}
{\bibfield  {journal} {\bibinfo  {journal} {Phys. Rev. A}\ }\textbf {\bibinfo
		{volume} {87}},\ \bibinfo {pages} {022310} (\bibinfo {year}
	{2013})}\BibitemShut {NoStop}%
\bibitem [{\citenamefont {Arrighia}\ and\ \citenamefont {Patricot}(2004)}]{AP}%
\BibitemOpen
\bibfield  {author} {\bibinfo {author} {\bibfnamefont {P.}~\bibnamefont
		{Arrighia}}\ and\ \bibinfo {author} {\bibfnamefont {C.}~\bibnamefont
		{Patricot}},\ }\href@noop {} {\bibfield  {journal} {\bibinfo  {journal} {Ann.
			Phys.}\ }\textbf {\bibinfo {volume} {311}},\ \bibinfo {pages} {26} (\bibinfo
	{year} {2004})}\BibitemShut {NoStop}%
\bibitem [{\citenamefont {Duvenhage}\ and\ \citenamefont {Snyman}(2018)}]{DS}%
\BibitemOpen
\bibfield  {author} {\bibinfo {author} {\bibfnamefont {R.}~\bibnamefont
		{Duvenhage}}\ and\ \bibinfo {author} {\bibfnamefont {M.}~\bibnamefont
		{Snyman}},\ }\href@noop {} {\bibfield  {journal} {\bibinfo  {journal} {Ann.
			Henri Poincar{\'e}}\ }\textbf {\bibinfo {volume} {19}},\ \bibinfo {pages}
	{1747} (\bibinfo {year} {2018})}\BibitemShut {NoStop}%
\bibitem [{\citenamefont {Schwinger}(1960)}]{S}%
\BibitemOpen
\bibfield  {author} {\bibinfo {author} {\bibfnamefont {J.}~\bibnamefont
		{Schwinger}},\ }\href@noop {} {\bibfield  {journal} {\bibinfo  {journal}
		{Proc. Nat. Acad. Sci. U.S.A.}\ }\textbf {\bibinfo {volume} {46}},\ \bibinfo
	{pages} {570} (\bibinfo {year} {1960})}\BibitemShut {NoStop}%
\bibitem [{\citenamefont {Ruzzi}\ \emph {et~al.}(2005)\citenamefont {Ruzzi},
	\citenamefont {Marchiolli},\ and\ \citenamefont {Galetti}}]{RMG}%
\BibitemOpen
\bibfield  {author} {\bibinfo {author} {\bibfnamefont {M.}~\bibnamefont
		{Ruzzi}}, \bibinfo {author} {\bibfnamefont {M.~A.}\ \bibnamefont
		{Marchiolli}},\ and\ \bibinfo {author} {\bibfnamefont {D.}~\bibnamefont
		{Galetti}},\ }\href@noop {} {\bibfield  {journal} {\bibinfo  {journal} {J.
			Phys. A}\ }\textbf {\bibinfo {volume} {38}},\ \bibinfo {pages} {6239}
	(\bibinfo {year} {2005})}\BibitemShut {NoStop}%
\bibitem [{\citenamefont {Marchiolli}\ \emph {et~al.}(2005)\citenamefont
	{Marchiolli}, \citenamefont {Ruzzi},\ and\ \citenamefont {Galetti}}]{MRG}%
\BibitemOpen
\bibfield  {author} {\bibinfo {author} {\bibfnamefont {M.~A.}\ \bibnamefont
		{Marchiolli}}, \bibinfo {author} {\bibfnamefont {M.}~\bibnamefont {Ruzzi}},\
	and\ \bibinfo {author} {\bibfnamefont {D.}~\bibnamefont {Galetti}},\
}\href@noop {} {\bibfield  {journal} {\bibinfo  {journal} {Phys. Rev. A}\
	}\textbf {\bibinfo {volume} {72}},\ \bibinfo {pages} {042308} (\bibinfo
	{year} {2005})}\BibitemShut {NoStop}%
\bibitem [{\citenamefont {Marchiolli}\ \emph {et~al.}(2009)\citenamefont
	{Marchiolli}, \citenamefont {Silva},\ and\ \citenamefont {Galetti}}]{MSG}%
\BibitemOpen
\bibfield  {author} {\bibinfo {author} {\bibfnamefont {M.~A.}\ \bibnamefont
		{Marchiolli}}, \bibinfo {author} {\bibfnamefont {E.~C.}\ \bibnamefont
		{Silva}},\ and\ \bibinfo {author} {\bibfnamefont {D.}~\bibnamefont
		{Galetti}},\ }\href@noop {} {\bibfield  {journal} {\bibinfo  {journal} {Phys.
			Rev. A}\ }\textbf {\bibinfo {volume} {79}},\ \bibinfo {pages} {022114}
	(\bibinfo {year} {2009})}\BibitemShut {NoStop}%
\end{thebibliography}
\end{document}